\documentclass[10pt,a4paper,final]{article}
\usepackage[a4paper, total={6in, 8.5in}]{geometry}
\usepackage[utf8]{inputenc}
\usepackage{amsmath}
\usepackage{amsfonts}
\usepackage{amssymb}
\usepackage{authblk}
\usepackage{graphicx,wrapfig}
\usepackage{hyperref}
\title{{\bf NMR studies of electronic properties of solids}}
\author{Henri Alloul \thanks{Henri Alloul (2014), {\it NMR studies of electronic properties of solids}, Scholarpedia, 9(9):32069.}}
\affil{LPS - CNRS/Universite Paris Sud, Orsay, France}
\date{}

\begin{document}
\maketitle
\begin{abstract}
The NMR technique is quite essential as it permits atomic scale measurements in materials. The aim of this article is to present the main parameters accessible to NMR experiments and to relate them to the electronic properties of simple solids in which electronic correlations are not dominant. It will be shown that in metals the Pauli paramagnetism, that is the electronic density of states is accessible through NMR shift and spin lattice relaxation data. Those permit as well to determine electronic gaps in some electronic structures, or in the SC state. The capability to reveal by NMR the incidence of impurities and disorder is also underlined. A  complementary article will be dedicated to the specific properties of strongly correlated materials which have been revealed and studied by NMR experiments.
\end{abstract}

\section{Introduction}
\label{Introduction}
The electronic properties of solids were on the first half of the twentieth century considered mostly in the frame of an independent electron approximation with spin degeneracy. Their resulting electronic band structure is such that each electronic level could be doubly occupied. In such an approach one expects metals or insulators with no significant
magnetic properties. 

The traditional experimental studies of the electronic states in such solids usually require a determination of their electronic band structure and how it affects the physical properties. Those are usually determined by measurements taken at the macroscopic scale such as optical, transport and magnetic data. To go beyond these approaches the Nuclear Magnetic Resonance is thoroughly used. This technique, discovered in the mid of the 20th century is quite essential as it permits to perform atomic scale measurements in the materials, differentiating the properties which can be attributed to specific phases or sites in the structure. 

Let us recall that the physical properties of solids, and in particular their magnetic properties, are determined by the electronic states. On the other hand, the nuclear spin moments, which do not affect these properties, provide an extremely useful probe for the electronic properties. Atomic nuclei are made up of neutrons and protons, which are spin 1/2 particles. They are assembled into quantum states in which the nuclear ground state has a total spin $\vec{I}$ that may be integer or half-integer. The associated magnetic moment ${\bf \mu }_{\mathrm{n}}$ is proportional to the magnetic moment of the proton $\mu _{\mathrm{p}}$, where the multiplicative factor is analogous to the Land\'{e} factor for an atomic electronic moment. Each atomic nucleus thus has a specific magnetic moment ${\bf \mu }_{\mathrm{n}}=\hbar \gamma _{\mathrm{n}}\vec{I}$. The
gyromagnetic ratio $\gamma _{\mathrm{n}}$ is known to great accuracy for
each of the stable isotopes in the periodic table.

Since $\mu _{\mathrm{p}}\simeq 10^{-3}\mu _{\mathrm{B}}$, the nuclear moments are extremely small, as are their mutual interactions. As a consequence, they are almost always in a paramagnetic state with a Curie
magnetization ${\bf \mu _{z}=N(\hbar {\gamma _\mathrm{n} })^2 I(I+1)}B_{0}/3k_{B}T$.  We see immediately that, in a given applied field, the nuclear magnetisation is about $10^{6}$ times smaller than the electronic magnetisation. The nuclear magnetism is practically impossible to detect using just macroscopic magnetization measurements taken on bulk samples. But, although the nuclear magnetic susceptibilities are weak, they can be detected by magnetic resonance which is a spectroscopy that permits a selective detection of the
nuclear spin response. The weakness of the interaction between electronic and nuclear spins permits to consider the nuclear spins as somewhat ideal probes of the electronic properties of materials.

As the NMR technique is described at length in various articles in Wikipedia and Scholarpedia for its applications in chemistry, biology and medical sciences, we shall only recall below very briefly in section \ref{Mag_res} the technical principles of NMR, but will describe in more detail in section \ref{Elec_hfine_coupling} the couplings between nuclear spins and electron spins. Those allow one to probe the electronic properties of solids and their magnetic response through the determination of the shift of the NMR signal (section\ref{NMR shifts}). In section \ref{Spin lattice relaxation}, we recall the spin lattice relaxation processes which permit to use the dynamics of the nuclear magnetisation to probe the excited electronic states in condensed matter.  The pairing of electrons in the superconducting state of simple metals also yield remarkable effects on the NMR (section \ref{NMR_super}). We shall recall then in section \ref{Quadrupolar} that while spin $1/2$ nuclei only probe electronic spin excitations, nuclear spins probes with $I>1/2$ sense as well an electrostatic quadrupole interaction which affects the nuclear spin levels and gives access to charge ordering in materials. Finally, in section \ref{Impurities}, we evidence the local effects induced by impurities or disorder on the NMR spectra.

\section{The magnetic resonance phenomenon and some of its basic applications}
\label{Mag_res}
Discovered at the outset of the second world War by F. Bloch and E.M. Purcell, the nuclear magnetic
resonance technique has become quite immediately a unique method to investigate the chemical and physical properties of condensed matter (Abragam ,1961, Slichter ,1963). Its success results from the fact that it resolves spectroscopicaly the properties of the nuclear spins of the distinct atomic species present in
materials. A homogeneous applied external magnetic field $B_{0}$ induces a splitting of the nuclear spin energy level $h\nu _{L}=\hbar \gamma _{\mathrm{n}}B_{0}$ which usually falls in the radio-frequency range of the order of 10 MHz per Tesla. The absorption of a radio-frequency field at the adequate frequency $\nu _{L}$ permits to detect the presence of the corresponding nuclei.

One highlight of NMR is the acquired possibility to provide images of the spatial distribution of $^{1}$H nuclei in-vivo in biological matter, which is the basis of medical Magnetic Resonance Imaging. Though this is the popular application of NMR known from a large audience, NMR is an even more powerful technique when one uses the interactions of given nuclear spins in a material with their neighboring atomic states. This results in rich spectroscopic splittings of the NMR lines which permit to locate the atoms in molecular states and therefore to determine the molecular structures in chemistry or in the solid state (''Nuclear Magnetic Resonance'' in Wikipedia). Such spectroscopic techniques have been revealed since the 1980s but their impact has been tremendously highlighted by the improvement of the SC magnet industry which has allowed to produce extremely homogeneous magnetic fields $B_{0}$ as large as 21 Tesla with negligible drift in time. In the corresponding range of frequencies $\nu _{L}$, exceptionally stable coherent sources are available, with narrower spectral widths than the transitions to be observed. These are obtained by electronic oscillators with frequency stabilized on the resonant mode of a piezoelectric quartz crystal. Moreover, for such frequencies, very powerful amplifiers are also available. With radio-frequency pulses, these can be used to significantly modify the populations of the spin quantum states. Such setups are useful for studying nuclear relaxation. Therefore on the electronics side, NMR has highly benefited of all the developments of fast semi-conductor and computing capabilities associated with the expansion of information technologies.

\section{ Electronic hyperfine couplings}
\label{Elec_hfine_coupling}
It is clear that changes in the magnetic induction in a material can be detected directly by a change in the nuclear Larmor frequency. In weakly magnetic materials, for which the magnetization is negligible, $B_{0}=\mu
_{0}H_{a}$ and the nuclear Larmor frequency $\nu _{L}$ should be determined solely by the applied external field $H_{a}$. It would be difficult to obtain information about the physical properties of materials in such a limit. But we have to recall that the nucleus is a kind of atomic scale microscopic probe, coupled to the electrons. Interactions like the dipole interactions between nuclear and electronic spins are such that the nuclear spin feels a magnetic field associated with the polarization of the electronic magnetic moments. This means that the magnetic field felt by the nuclear spins is modified with respect to the applied field. It is the spectroscopy of these fields that provides atomic scale information about the immediate vicinity of the nuclei in the material. Let us examine the different interactions between the nuclear spins and the magnetic moments of electronic origins, known collectively as hyperfine interactions (Abragam 1961, Slichter 1963).

The dipole interaction between the moments associated with a nuclear spin $\mathbf{I}$ and an electron spin $\mathbf{S}$ separated by a displacement $\mathbf{r}$ is

\begin{equation}
\label{eq:Hdd}
H_{dd}=-\frac{\mu_{0}}{4\pi }
\frac{\gamma _{e}\gamma _{n}\ \hbar ^{2}}{r^{3}}
\left[ \mathbf{I}.\mathbf{S}-\frac{3(\mathbf{I}.\mathbf{r})(\mathbf{S}.\mathbf{r})}{r^{2}}\right]
\end{equation}

where $\gamma _{n}$ and $\gamma _{e}$ are the nuclear and electronic gyromagnetic moments, respectively, and $\mathbf{S}$ and$\mathbf{\ I}$ are here dimensionless quantities. This dipole interaction diverges when $r$ tends to zero, and is only therefore valid for electrons with zero probability of being at the site of the nucleus. This is the case for electrons in the $p,d$, or $f$ shells. On the other hand, the$\ s$ electrons have nonzero probability of being at the site of the nucleus. The Dirac Hamiltonian can be used to show that the corresponding interaction, called the contact interaction $H_{c}$ is scalar, and is given in this case by
\begin{equation}
\label{eq:H Contact}
H_{c}=\frac{%
\mu
_{0}}{4\pi}8\pi ^{3}\ \gamma _{e}\gamma _{n}\ \hbar ^{2\ }\ \mathbf{%
I}.\mathbf{S\ }\delta (\mathbf{r}). 
\end{equation}

Finally, the interaction with the magnetic field associated with the orbital angular momentum of the electron is
\begin{equation}
\label{eq:Horb}
H_{orb}=-\frac{%
\mu
_{0}}{4\pi }\gamma _{e}\gamma _{n}\  \hbar ^{2\ }\ \frac{\mathbf{I}.%
\mathbf{\ell }}{r^{3}}\mathbf{\ }. 
\end{equation}

These Hamiltonians can all be written in the form
\begin{equation}
\label{eq:Heff}
H_{eff}=-\gamma _{n}\ \hbar \mathbf{\ \ I}.\mathbf{B}_{eff}, 
\end{equation}

and we may consider that each electron induces a magnetic field $\mathbf{B}_{eff}$ at the nuclear site. As the temporal fluctuations of the electronic moments are very fast compared with the nuclear Larmor frequency, the static component of $\mathbf{B}_{eff}$ is its time average. The position of the NMR for a given nucleus is thus determined by the time average $\mathbf{B}_{eff}$ of the resultant of the fields due to the different electrons in the material. It is easy to see that the hyperfine interaction will vanish for filled electronic shells, because they have zero total spin and total orbital angular momentum. When there is no applied field, $\mathbf{B}_{eff}$
can only be nonzero for materials in which there is a static spin or orbital magnetic moment. This will be the case for magnetically ordered materials.

\section{NMR shifts}
\label{NMR shifts}
In a material, each of the distinct hyperfine couplings listed above induces a specific contribution to the NMR shift. Those are of course quite dependent on the magnitude of the corresponding hyperfine coupling and of the electronic state of the considered material.
\subsection{Chemical shifts}

In substances where the electrons are paired in atomic or molecular orbitals, the static part of the hyperfine coupling is only nonzero in the presence of an applied field $\mathbf{B}_{0}$, and is proportional to $\mathbf{B}_{0}$, like the magnetization. The resonance is shifted with respect to that of the free atom in a gas. The relative shift $\mathbf{B}_{eff}/\mathbf{B}_{0}$ may be due to the orbital part of the hyperfine coupling. This is the case, for example, for the displacement due to the orbital currents induced by the external magnetic field in electronic or molecular shells close to the nucleus. Since this shift depends on the electronic charge distributions, it is highly sensitive to the chemical environment of the given atom, hence the name chemical shift. These effects are generally small, and expressed in in parts per million (ppm) but can be used to
distinguish the nuclear spin resonances of the different atoms depending on their environment. This has become a very powerful tool, used universally in chemistry and biology. Routine chemical analyses are carried out by NMR. It also helps one to determine the 3D structures of biological molecules, using multidimensional methods, which have reached an exceedingly high level of refinement.

\subsection{Knight shifts in Metals}

When the electron states are not paired in molecular states or in bonds, the spin degeneracy of the electronic states might be lifted by the applied field, as for electron states at the Fermi level in a metallic band. In that case a $\mathbf{B}_{eff}$ component due to the electronic atomic moment may arise via the contact hyperfine
term. In a metal the corresponding frequency shift $K$ of the Larmor frequency is called Knight shift (Knight 1967) and is directly proportional to the Pauli spin susceptibility $\chi _{P0}$ of the metallic band. In usual metallic systems such as alkali metals the early day studies by NMR permitted to demonstrate that this technique gives indeed the best evaluation of the electronic spin susceptibility. Assuming that the main hyperfine coupling is the direct on-site contact interaction of Equ(2), which can be written

\begin{equation}
H_{c} =A_{0}\mathbf{I.s}\delta(\mathbf{r}) 
\label{eq:Hc}
\end{equation}

this yields an NMR Knight shift

\begin{equation}
\label{eq:KSki}
K=A_{0}\chi _{P0}/\ (g\mu _{B}\hbar \gamma _{\mathrm{n}})
\end{equation}

In such simple metals the spin susceptibility $\chi _{P0}=(g\mu _{B})^{2}\rho (E_{F})/2$  measures the actual density of states $\rho (E_{F})$ taken per spin direction at the Fermi level, which is typically $T$ independent as the conduction electron bandwidth is usually quite large, and the Fermi level much higher than $k_{B}T$.
The Knight shift is usually a large quantity which is measured in \%. This comes about because the contact coupling $A_{0}$ is usually much larger than the corresponding dipole or orbital couplings, which permits to sense very effectively the Pauli susceptibility.

\subsection{NMR in Magnetic Materials}
In magnetic materials, the electronic moments are static at low temperatures, as compared with their behavior at the Curie or N\'{e}el temperatures. It follows that the static effective fields are nonzero even in the absence of any applied field. For atomic nuclei carrying an electronic moment, this field will be very large (several Tesla in general), and will give rise to a resonance at the Larmor frequency $h\nu _{L}=\hbar \gamma _{\mathrm{n}}B_{eff}$, which can be detected in the absence of an applied external field. One speaks then of Zero Field Nuclear Magnetic Resonance (ZFNMR).  The fields induced on the nuclear spins of non magnetic atomic sites are generally weaker but can still give valuable information on the properties of the magnetic state  as will be seen later on in various magnetic materials.

\subsection{1D metals}
The incidence of lattice dimensionality can give rather original effects, the one dimensional case (1D) being really an important one especially near half filling, that is with one electron per atom in atomic chains. One does usually expect a dimerisation of the chains at low $T$ which induces the opening of a gap at the Fermi level at low $T$ and a corresponding so called Peierls transition from a metal to an insulator (see ``Peierls transition'' in Wikipedia). This effect has been observed in many materials with linear metallic chains. One of the first evidence for such a Peierls transition has been found from Pt NMR data taken (Niedoba \textit{et. al.} 1973) in the Low-Temperature Insulating State in the One-Dimensional conductor K$_{2}$Pt(CN)$_{4}$Br$_{0.3}$, 3H$_{2}$O. A similar effect is seen in the case of the organic salt TTF-TCNQ, although that case is slightly more complicated as two successive ordering transitions occur for the two type of chains involved in that structure (Jerome and Schulz 2002). Let us point out that such a metal insulator transition often occurs solely because the elastic energy loss produced by the Peierls dimerisation is encompassed by the energy gain associated with the opening of the electronic gap in the band structure. This can be unrelated to the existence of electronic correlations. More important phenomena are expected to happen in metallic 1D systems in presence of strong electronic correlations.

\section{Spin lattice relaxation}
\label{Spin lattice relaxation}
The local susceptibility measurements are giving pertinent information on the electronic properties of the material in its ground state. But NMR also permits to probe the excited electronic states through the fluctuations of the local field $\mathbf{B}_{eff}$. This occurs through the nuclear spin lattice relaxation (NSLR) processes which drive back the nuclear spins towards their thermodynamic equilibrium once the latter has been disturbed intentionally.

Indeed the nuclear spin magnetization is not established immediately if an external magnetic field is applied instantaneously to the material. The very interactions between the nuclear spins and the electronic degrees of freedom govern the spin lattice relaxation time $T_{1}$ which is required to establish thermodynamic equilibrium. One typically needs transverse local field components at the Larmor frequency $h\nu _{L}$ to induce the difference of population of the nuclear spin levels. Therefore $T_{1}$ is directly linked with the transverse field Fourier component at $\nu _{L}$ of $\mathbf{B}_{eff}(t)$ . One can see that, for the hyperfine couplings considered above, this resumes in a measurement of the electronic dynamic susceptibility of the electron spin system. 

So, in systems with unpaired spins the dominant $T_{1}$ process is due to local field fluctuations induced by the dynamics of the local electronic magnetization. Theoretically, the spin contributions to $(T_{1}T)^{-1}$ may
be written using the imaginary part of the dynamical electron spin-susceptibility ${\chi ^{\prime \prime }(\mathbf{q},\nu _{n})}$ as 

\begin{equation}
(T_{1}T)^{-1}=\frac{2\gamma _{n}^{2}k_{B}}{g^{2}\mu _{B}^{2}}\sum_{\mathbf{q}%
}|A_0|^{2}\frac{\chi ^{\prime \prime }(\mathbf{q},\nu _{n})}{\nu
_{n}}.  \label{eq:1/T1T}
\end{equation}

Here, as for the Knight shift, we assumed that the dominant hyperfine coupling is the contact term. 
We shall see later many examples which permit to evidence that $T_{1}$ measurements permit one to monitor the occurrence of phase transitions, to give relevant informations on energy gaps between the ground state and excited states in many electronic systems. In cases where some ionic species are mobile in a material, as for instance in ionic conductors, the atomic diffusion processes can govern the local field fluctuations sensed on some nuclear spin sites, and the $T_{1}$ measurements may permit to monitor these ionic diffusion motions.

\subsection{Spin lattice relaxation in standard 3D metals}

For a simple metallic band, the  dynamic electronic susceptibility response is simple enough and one writes

\begin{equation}
[\label{eq:kiomega}{\chi }_{0}(\omega )=\sum\nolimits_{\mathbf{q}}{\chi }_{0}(\mathbf{q}),\omega )=\chi _{P0}\left[ 1+i\pi \hbar {\omega \ }\rho (E_{F})\right] .
\end{equation}

One can immediately see that this yields a simple expression for the spin lattice relaxation from Equ (7)

\begin{equation}
\label{eq:T1T}(T_{1}T)^{-1}=\pi k_{B}\ A_{0}^{2}\ \rho ^{2}(E_{F})/\hbar 
\end{equation}

so that a universal relation holds between the Knight shift and $T_{1}$.

\begin{equation}
\label{eq:Korringa}K^{2}T_{1}T=\mathcal{S=}\left( \hbar /4\pi k_{B}\right) \left( g\mu
_{B}/\hbar\gamma _{\mathrm{n}}\right) ^{2}. 
\end{equation}

\begin{figure}
\begin{center}
\includegraphics[height=6cm,width=7.5cm]{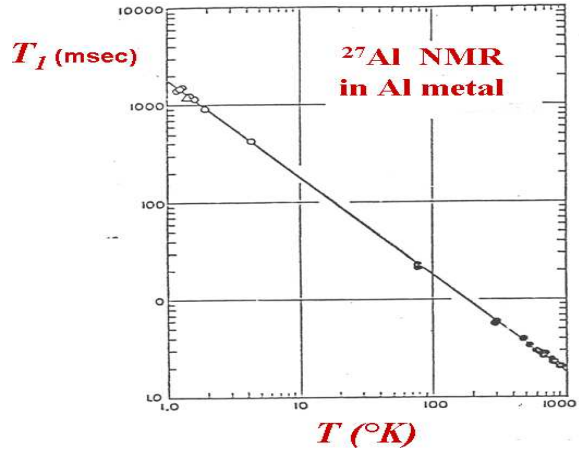} 
\caption{\label{fig:Fig1}The $^{27}$Al spin lattice relaxation time $T_{1}$  measured in pure aluminium metal is plotted versus temperature in a log/log scale. The linear fit represents the relation $T_{1}T =$1.85 sec $^{\circ}K$.} 
\end{center}
\end{figure}

As $K$ is $T$ independent, this so called "Korringa" relation applies rather well in the absence of electronic correlations.  As an example one can see in Fig.~\ref{fig:Fig1} that $T_{1}T$ of $^{27}$Al is constant in pure aluminium metal on a $T$ range which extends over more than three orders of magnitude. The $T_{1}$ value in a metal is quite often used to define an empirical temperature scale especially in the very low $T$ regime below 1 K.

\subsection{Incidence of weak electron correlations}

So far we did not consider any influence of electronic correlations though even in simple alkali metals electron-electron interactions play a role in the electronic scattering processes. We also do know that some electronic
systems are on the verge to become magnetic. Those quasi AF or quasi ferromagnetic metals can be identified by the very fact that the Korringa relation does not apply then straightforwardly as the spin susceptibility
does not behave as described above for free electron Fermi liquid systems. In such cases the dynamic spin susceptibility is not uniform in $\mathbf{q}$ space as was shown initially by (Moriya 1956), and exhibits enhanced values either for $\mathbf{q}=0$ for nearly ferromagnetic metals or for an AF wave vector $%
\mathbf{q}_{\mathbf{AF}}\ $for nearly AF materials. 

In the former case the $\mathbf{q}=0$ spin susceptibility is enhanced by a factor ${S}=1/(1-{I}\chi_{P0})$ usually called the Stoner factor, and $\chi_{P}=S \chi_{P0}$ exhibits a large increase with decreasing $T$ and only saturates at very low $T$, as has been illustrated in the nearly ferromagnetic metals like TiBe$_{2}$  or in
elemental Pd metal. In those cases, the enhancement of the dynamic spin susceptibility $\chi_{0}(\mathbf{q}\omega)$ is not uniform in $\mathbf{q\ }$ space and is weaker for $\mathbf{q}\neq 0$, therefore the Knight shift is more enhanced than $(T_{1}T)^{-1}$. The Korringa relation does only apply when $\chi _{P}\ $saturates (Alloul and Mihaly 1982), with a $T=0$ Korringa constant $ K^{2}T_{1}T=\Delta \ \mathcal{S}$ increased by an ${S}$ dependent factor $\Delta({S})$. On the contrary in nearly AF metals $\chi_{0}(\omega)$ is peaked
for $\mathbf{q}= \mathbf{q}_\mathbf{AF}$ which means that the static spin susceptibility and the Knight shift $K $ are less unhanced than $\chi_{0}(\mathbf{q}_\mathbf{AF},\omega)$. Correspondingly $K^{2}T_{1}T=\Delta \ \mathcal{S}$ corresponds in that case to$\ \Delta <1$, that is a decreased Korringa constant, as has been seen for instance in the compound MnSi (Corti et al 2007).

\section{ NMR in superconductors}
\label{NMR_super}

Obviously, the establishment of a SC state yields profound transformations of the electronic properties which will be seen in the NMR response. NMR experiments do not only evidence the occurrence of SC. They also permit to characterize the properties of the SC electronic state (Mac Laughlin 1976).  
 
\subsection{Knight shift, relaxation and gap in the SC state}

\begin{figure}
\begin{center}
\includegraphics[height=6cm,width=7.5cm]{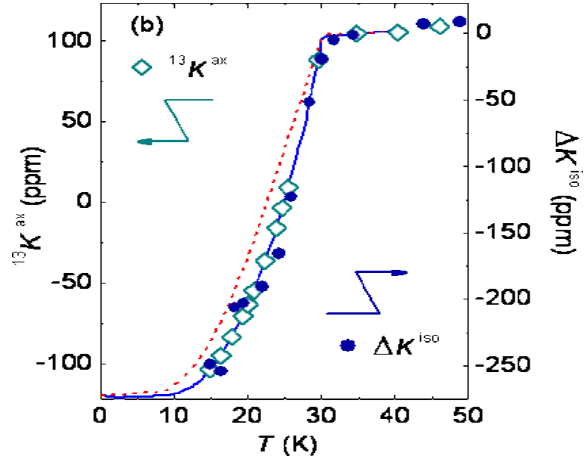}
\caption{\label{fig:Fig2}The $^{155}$Cs and $^{13}$C NMR shifts measured in the Cs$_{3}$C$_{60}$ phase are plotted versus $T$ below  the superconducting temperature $T_{c} =30 $K. The NMR shifts follow the standard Yosida type decrease expected for singlet superconductivity (Wzietek {\it et al.} 2014).} 
\end{center}
\end{figure}

One of the major effects which occur for phonon mediated SC in usual metals is the pairing of electrons in a singlet state. Such a pairing suppresses totally the normal state spin susceptibility at $T=0$. This is seen quite simply as a full suppression of the spin contribution $K_{s}$ to the Knight shift in NMR.  In type I superconductors, the magnetic induction vanishes in the Meissner state which by itself forbids observation of the NMR signal in the SC state. But in type II superconductors the field penetrates as an array of vortices which becomes so dense near the upper critical field  $H_{c2}$ that it becomes possible to detect the NMR signal in that regime, and to see the suppression of the Knight shift. Taking into account the variation of the SC gap and the thermal population at temperatures near $T_{c}$, yields a specific $T$ dependence of the spin susceptibility, that is of $K_{s}(T)$, which has been computed by (Yosida 1956), and which is given by

\begin{equation}
K_{s}(T)/K_{n}=\int_\Delta^\infty[N(0)|E|/(E^2-\Delta^2)^{1/2}](df/dE)dE
\label{eq:Yosida}
\end{equation}
Here $f$ is the Fermi function. The actual variation of the knight shift with decreasing temperature can be measured and displays an agreement with this Yosida function as can be seen in Fig.~\ref{fig:Fig2}.

As for spin lattice relaxation data, it can be taken in type I SC using ingenious tricks such as experiments in which the external field is cycled from a field exceeding the critical field $H_{c}$ down to a field $H<H_{c}$ in which the nuclear spin magnetization is let free to evolve under the influence of the electronic system.

\begin{figure}
\begin{center}
\includegraphics[height=6cm,width=7.5cm]{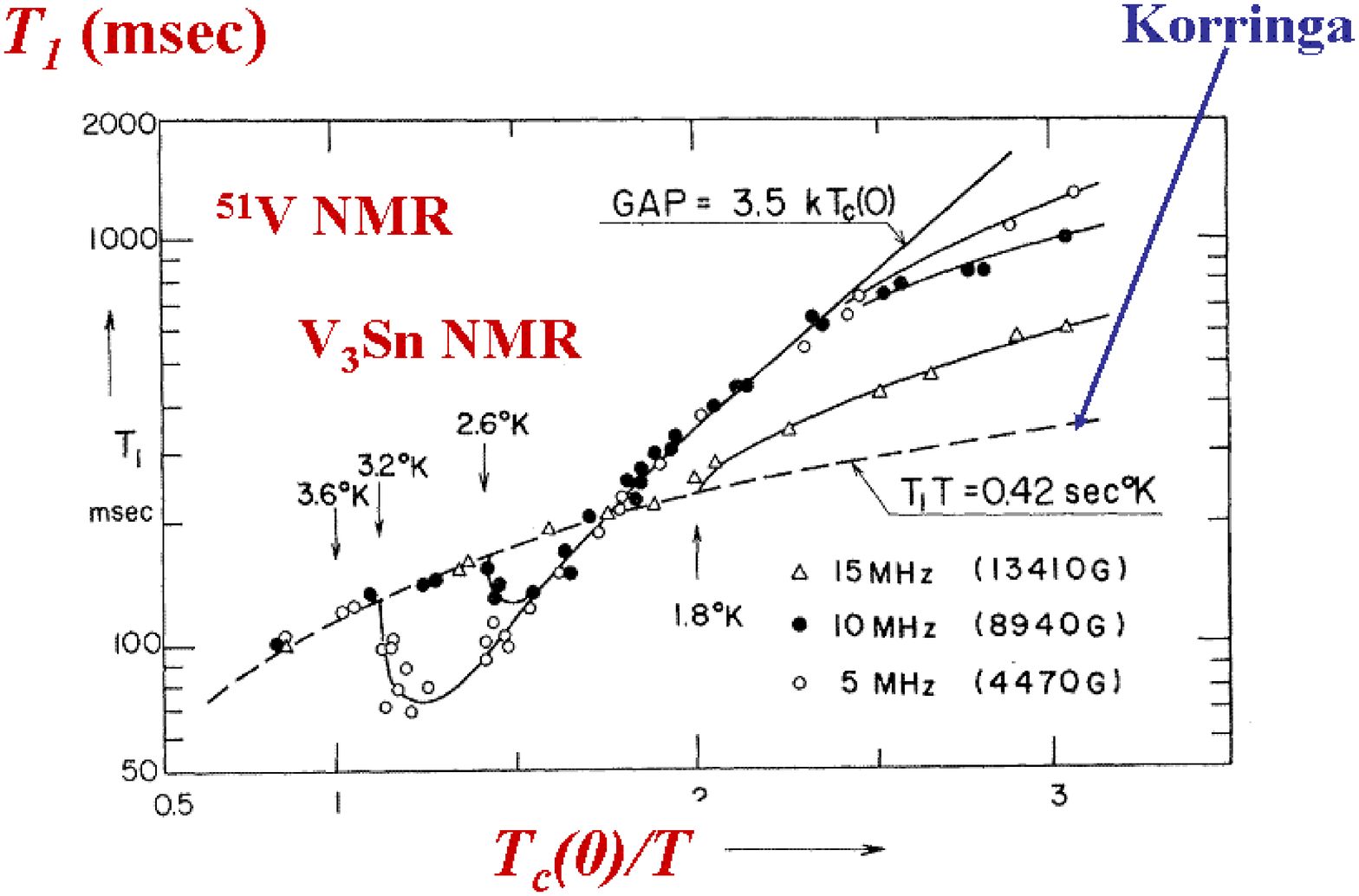}
\caption{\label{fig:Fig3}The $Log(1/T_{1})$ of $^{51}$V in $V_{3}$Sn is plotted versus $1/T$ for three distinct applied fields, which induce changes of $T_{c}$. In the normal state above $T_{c}$ the relaxation rate is field independent with $T_{1}T$=0.42  sec$^{\circ}K$. Below $T_{c}$ the reduction of $T_{1}$ represents the Hebel Slichter coherence peak. At low $T$ all curves point towards an activated behavior associated with the full opening of the superconducting gap (adapted from Mac Laughlin 1976).} 
\end{center}
\end{figure}

The opening of the SC gap yields an activated exponential increase of the
NMR spin lattice relaxation rate as $T$ approaches $0$ , which permits a determination of the SC gap magnitude from the corresponding low $T$ variation of $T_{1}^{-1}$ (Fig.~\ref{fig:Fig3}). However the great advance of BCS theory has been its ability to describe the excited states in the SC state up to $T_{c}$. Indeed in such a BCS SC state subtle effects are revealed by $T_{1}T$ data taken near $T_{c}$. An increase of the spin lattice relaxation rate above the normal state Korringa value takes place below $T_{c}$. This so called  coherence peak (Hebel and Slichter, 1957) results partly from the thermal
population of the increased density of electronic states which piles up above the SC gap. The $T_{1}$ only lengthens at somewhat lower temperatures than $T_{c}$ (see Fig.~\ref{fig:Fig3}).

Both these discoveries of the decrease of the spin susceptibility and of the occurrence of a Hebel-Slichter coherence peak have given the early evidences for the applicability of BCS theory of superconductivity in usual metallic systems.

\subsection{Field distribution in the mixed state of type II superconductors}

In type II superconductors the magnetic induction varies significantly in space in the mixed state. This leads to a distribution of Larmor frequencies for the nuclear spins in the material. The shape of the NMR spectrum
reconstitutes the histogram of the magnetic fields. Close to $H_{c2}$, singularities appear in the spectrum for values of the magnetic field corresponding to the extrema of the field distribution. The shape and  width of the observed resonance can be used to deduce $\lambda$, the magnetic field penetration depth. However, for
experimental reasons, NMR is not the best method for studying the superconducting state. A related technique uses elementary particles called muons. These behave like heavy electrons (or light protons), and have the
property of decaying by emission of positrons in the direction of their spin. A muon whose spin is initially polarised perpendicularly to the field $B_{0}$ is implanted in the sample at time zero. One then observes the direction of the emitted positrons when it decays. By repeating this experiment for a large number of events, the free precession signal of the muon spin can be reconstructed statistically. This experiment is equivalent to an NMR experiment, and can be used to determine $\lambda$. Since muons can be implanted in almost any sample, it has been possible to make comparative measurements of $\lambda$ in a wide range of superconducting
materials.

\section{ Quadrupolar electrostatic interactions and NQR}
\label{Quadrupolar}
Generally in solid state NMR an atomic nucleus with spin $\mathbf{I}$ has its spin energy levels determined by the Zeeman interaction with the external magnetic field $\mathbf{B}_{0}$. But nuclei with $I>1/2$ , in
which the nuclear charge has a non spherical shape are also characterized by the quadrupole moment $Q$ of charges . This charge distribution interacts with the electrostatic electric fields (in fact the electric field gradient
EFG) induced on the nucleus by the distribution of electronic charges in the material. Therefore the nuclear spins are influenced both by the Zeeman and the quadrupolar Hamiltonian $\mathcal{H}_{Q}$, which can be
written  (Abragam 1961, Slichter 1963)

\begin{equation}
\label{eq:HQ} 
\mathcal{H}_Q = \frac{eQ}{2l(I-1)}\sum_{\alpha }V_{\alpha \alpha }I_{\alpha}^2
\end{equation}

where $V_{\alpha \alpha }$ are the second derivatives of the electric potential, that is the EFG tensor which has $\alpha =X,Y,Z$ as principal axes. The three quadrupolar frequencies 

\begin{equation}
\label{eq:nuQ}
\nu _{\alpha }=\frac{3eQV_{\alpha \alpha }}{ 2I(2I-1)h}
\end{equation}

are linked by Laplace equation $\sum_{\alpha }V_{\alpha \alpha}=\sum_{\alpha }\nu _{\alpha }=0$. Therefore it is more common to use two parameters - the quadrupolar frequency $\nu _{Q}=\nu _{Z}$ corresponding to
the largest principal axis component $V_{ZZ}$ of the EFG tensor and the asymmetry parameter $\eta =(V_{XX}-V_{YY})/V_{ZZ}$ (here the principal axes of the EFG tensor are chosen following $\left\vert V_{ZZ}\right\vert \geq \left\vert V_{YY}\right\vert \geq \left\vert V_{XX}\right\vert $).

\subsection{ Quadrupole split NMR spectrum}

\begin{figure}
\begin{center}
\includegraphics[height=8cm,width=8cm]{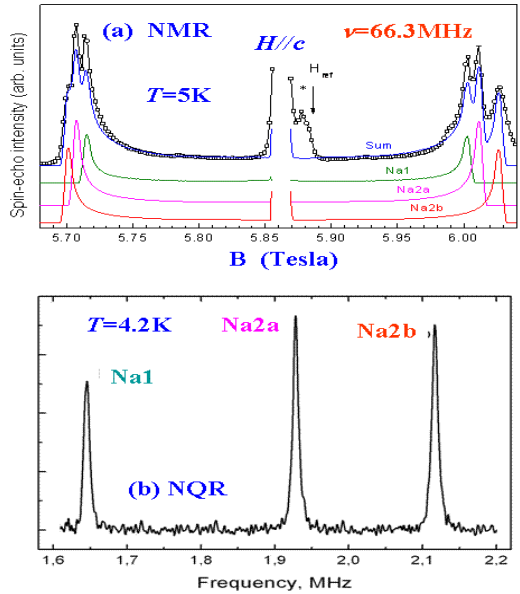}
\caption{\label{fig:Fig4}The quadrupole split NMR spectrum of $Na_{2/3}CoO_2$ is composed of three triplet spectra associated with three different Na sites (Alloul et al 2009). (b) The NQR spectrum of the same compound displays three independent NQR lines. (Platova et al 2009).} 
\end{center}
\end{figure}

The quadrupolar term modifies the NMR spectrum and can be treated as a perturbation if the quadrupole term is small as compared to the Zeeman splitting. Let us consider as an example the $^{23}$Na NMR in the layered cobaltates Na$_{x}$CoO$_{2}$. The $^{23}$Na nuclear spin is $I=3/2$ and therefore for a given direction of applied magnetic field $H_{0}$ relative to the crystal the NMR spectrum for a single Na site consists of 3 lines - a central line which corresponds to the $-1/2\leftrightarrow 1/2$ transition and two satellites corresponding to the $-3/2\leftrightarrow -1/2$ and $1/2\leftrightarrow 3/2\ $transitions. The position of the central line is determined by the applied field and the distance between the two satellite lines depends on the orientation of the external field with respect to the principal axis of EFG tensor described by the spherical
angular coordinates $\theta $ and $\varphi $ which can be expressed as 

\begin{equation}
\Delta \nu =\nu _{Q}(3\cos ^{2}\theta -1+\eta \sin ^{2}\theta \cos 2\varphi).  
\label{eq:SatelDist}
\end{equation}

In Fig.~\ref{fig:Fig4}(a), we show such a spectrum obtained with the field $B$ perpendicular to the CoO$_{2}$ layers in Na$_{x}$CoO$_{2}$ for $x=2/3$. One can see that the spectrum is somewhat complicated and can
be fitted by three spectra with different $\nu _{Q}$ that is three different Na sites. Such a spectrum permits to establish that the Na sites are in an ordered pattern with three sites per unit cell for this Na composition (Alloul et al 2009).

\subsection{Pure quadrupole resonance : NQR}

One can immediately see that in the absence of an applied external field the quadrupole Hamiltonian lifts partially the degeneracy of the spin states. For instance for a spin $I=3/2$ one gets two levels which are separated in frequency  by  (Abragam 1961)  

\begin{equation}
\nu =\nu _{Q}\sqrt{1+\eta ^{2}/3} \label{eq:nuNQR}
\end{equation}

so that the quadrupole frequency can be determined directly in a pure nuclear quadrupole resonance experiment taken in zero magnetic field. It is for instance illustrative to see that the $^{23}$Na NQR of Na$_{2/3}$CoO$_{2}$ shown in Fig.~\ref{fig:Fig4}(b) indeed displays the three frequencies expected from the
analysis of the NMR spectrum, which perfectly confirms the existence of the three Na lattice sites in the unit cell (Platova et al 2009).

This example permits us to evidence that the nuclear spin resonance spectra of nuclei with $I>1/2$ contain information relevant to the atomic structure or electronic charge distributions in materials. Thus the quadrupole
parameters may be regarded as complementary tools with respect to diffraction techniques for investigating structural changes or charge ordering in solids. Let us point out that the NQR, being a zero magnetic field experiment, is quite useful in the studies of electronic properties in the SC state. This permits for instance to restrict the technical limitations in measurements of the spin lattice relaxation in type I superconductors.

\section{Impurities and disorder}
\label{Impurities}
All real crystalline materials contain structural defects. Those are often impurity atoms substituted to some atoms in the ideal structure, or disorder induced by vacancies on some sites of the atomic structure or by deviations to the ideal structural arrangement of atoms. The incidence of specific substituted impurities on the physical properties of the material is sometimes well understood. The essence of the observed phenomena is that an impurity is a local screened Coulomb potential, which ideally is a uniform perturbation in {\bf q} space, inducing a response which is inhomogeneous in real space but which reflects the response to all {\bf q} values. 
In a classical metallic system, since the response is homogeneous up to $\left\vert \mathbf{q}\right\vert =k_{F}$, the main detected feature comes from this truncation of the response at $k_{F}$, which yields the well known Friedel oscillations in the local density of states for a charge defect, and the RKKY oscillation for a spin defect. Such effects differentiate the electronic response on sites in the vicinity of the defect and can be detected by NMR experiments. We shall first illustrate this in the case of magnetic defects in sp metals and show that NMR spectra permit to probe directly the RKKY  oscillations. 

\subsection{ Local RKKY  spin density oscillations induced by magnetic impurities in metals}

If one substitutes a magnetic impurity on a lattice site of a metal such as Cu, the neighbouring sites are differentiated and the magnetic response depends on the distance to the impurity. For local moment impurities, Yosida (1956) calculated explicitly the spin density oscillations assuming that the free electron spin $s$ and local moment $S$ interact by an exchange interaction

\begin{equation}
H=-J\ S.\ s\ \delta (r). 
\label{eq:Exchange}
\end{equation}

\begin{figure}
\begin{center}
\includegraphics[height=6cm,width=8.5cm]{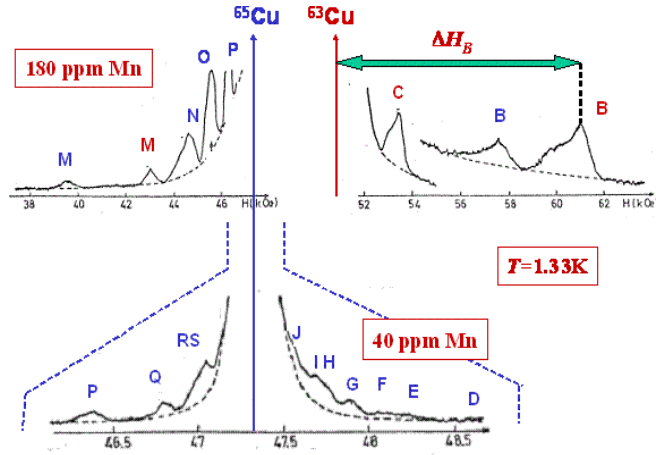}
\caption{\label{fig:Fig5}NMR spectra of the $^{63}$Cu and $^{65}$Cu nuclear spins in dilute Cu-Mn alloys obtained by sweeping the applied external field, at 1.3K. The spectra have been expanded vertically to exhibit the satellite lines by cutting the large intensity central lines at the pure copper NMR positions, which are pointed by arrows. On both sides of the central lines one can see the weak NMR signals of the diverse copper shells of neighbors. The sites far from the impurities are better resolved by reducing the impurity concentration, as shown in the expanded bottom spectrum (adapted from Alloul,2012).}
\end{center} 
\end{figure}

The resulting local spin density, calculated in perturbation theory at a position $R_{n}$ with respect to the impurity is given by 

\begin{equation}
n(R_{n})=-\frac{1}{4\pi }J\rho (E_{F})\ \frac{cos(2k_{F}R_{n})}{R_{n}^{3}}%
<S_{z}>, \label{eq:RKKY}
\end{equation}

for a field applied in the $z$ direction. For local moment impurities
such as Mn in Cu the NMR Knight shift of a Cu nuclear spin at position $%
R_{n} $ with respect to the impurity acquires an extra shift $\Delta K$
given in an applied field $H$ by

\begin{equation}
H\Delta K(R_n)=A_{hf}\ n(R_{n})=A(R_{n})<S_{z}>, 
\label{eq:Shift(R)}
\end{equation}
where $A_{hf}$ is the on site Cu hyperfine coupling. In most dilute alloys of transition elements only a few near-neighbor shells of the impurity could be resolved. However in Cu-Mn, the impurity magnetization $<S_{z}>$ becomes so large at low $T$ and high fields, that $n(R_{n})$ becomes sizable on many neighboring sites of the impurity. Up to 17 distinct shells of neighbors could be detected in that case, as can be
seen in the spectra of Fig.~\ref{fig:Fig5}, which gives a straightforward illustration of the occurrence of spin density oscillations. One can see that there are about as many extra lines (we call them satellite lines) on the right and on the left of the central line. The technical details about the assignment of the different lines to specific shells of neighbors, and the analysis of the spatial dependence of the spin polarization which can be deduced from those data are summarized in (Alloul 2012). One could push the analysis at a stage permitting to confirm
the overall $R_{n}^{-3}$ dependence, but also to evidence deviations from the asymptotic limit at short distances. Such deviations with respect to the spatial dependence of (17) could be analyzed by using a more reliable model than an exchange coupling between the impurity and electron spins.

\subsection{Transferred hyperfine couplings}

So far we have only considered hyperfine couplings between nuclear spins and the electrons involved in the atomic orbitals of the corresponding site. However the illustration of the RKKY interaction given in the section above did allow us to demonstrate that electrons on a given atomic site interact as well with the neighboring sites. Indeed if we consider Equ.(18) we see that it permits to define a transferred hyperfine coupling $A(R_{n})$ between the local moment at the origin and the nuclear spin at $R_{n}$.

Therefore in systems where different atomic sites are involved in a unit cell such transferred hyperfine couplings do play an important role. In particular, if one considers atomic sites which are not displaying the most important magnetic response they will still sense the response of the magnetic sites through such transferred hyperfine couplings.
In the case of Cu-Mn the transferred hyperfine coupling with the Mn magnetism extends on a large number of Cu sites. But, as the transferred hyperfine coupling decreases strongly with distance, it is often sufficient to consider solely transferred hyperfine couplings with the first nearest neighbors. For instance in a square lattice such as that of the CuO$_{2}$ planes of cuprate superconductors for which the electronic magnetic response is located on the Cu sites, the $^{63}$Cu nuclei will be coupled to the on site magnetic response and with that of the near neighbors. This can be cast in a combined wave-vector$\ \mathbf{q}$
dependence of the hyperfine coupling 
$A(\mathbf{q})=A_{0}+\sum A_{i}e^{i%
\mathbf{q}\mathbf{r}_{i}}$
in which $A_{0}$ is the on site local hyperfine
interaction between the observed nuclear and electron spins and $A_{i}$
is the hyperfine interaction with electron spins at neighboring sites at $
\mathbf{r}_{i}$. Similarly one has to substitute $A_{0}$ by $A(\mathbf{q})$ in the Equ (7) for the spin lattice relaxation (Moriya 1956).

This modification is important as it shows that the NMR experiments permit to sense the $\ \mathbf{q}$ dependence of the dynamic susceptibility in complex materials involving diverse atomic sites or a multiband
electronic structure. Combining NMR shift and macroscopic susceptibility data gives access to an experimental determination of $A_{i}$ which contains information on the electronic structure.

\subsection{ Disorder as viewed from quadrupolar effects}

We have seen in sec.\ref{Quadrupolar} that the nuclear spin resonance spectra of nuclei with $I>1/2$ contain information relevant to the atomic structure or electronic charge distributions in materials. Thus the quadrupole
parameters may be regarded as complementary tools with respect to diffraction techniques for investigating structural changes or charge ordering in solids. The NMR and NQR techniques are also important for
instance to characterize the local disorder in solids whereas the interpretation of the data on disordered materials from usual scattering experiments, such as X-ray or neutron scattering, is complicated by the
absence of the translation symmetry of long-range order. One will have the opportunity to discuss these aspects in relation with the physical study of correlated electron systems, which, being complicated materials, do involve many such defects.

\section*{References}
\begin{enumerate}
\item A. Abragam, ``The Principles of nuclear magnetism'' (Oxford: Clarendon Press, London, 1961).
\item Alloul H and Mihaly L . ``Ti NMR study of the nearly ferromagnetic system TiBe2''. Phys.Rev.Letters, 48, 1420 (1982).
\item H. Alloul, I. R. Mukhamedshin, T. A. Platova and A. V. Dooglav , ``Na ordering imprints a metallic kagom\'{e} lattice onto the Co planes of Na2/3CoO2'', Europhysics Letters 85 , 47006 (2009).
\item H. Alloul, ``From Friedel oscillations and Kondo effect to the pseudogap in cuprates", in J Supercond. Nov. Mag. 25, 385 (2012) ; (DOI) 10.1007/s10948-012-1472-x, arXiv:1204.3804.

\item M. Corti, F. Carbone, M. Filibian, Th. Jarlborg, A. A. Nugroho, and P. Carretta , ``Spin dynamics in a weakly itinerant magnet from Si29 NMR in MnSi'', Phys. Rev. B 75, 115111 (2007)

\item Hebel L.C. and Slichter C.P, ``Nuclear Spin Relaxation in Normal and Superconducting Aluminum'', Phys Rev 113, 1504 (1957)

\item Jerome D. and Schulz H. J.,``Organic conductors and superconductors'', Adv. Phys., 51, 293 (2002).DOI: 10.1080/00018730110116362

\item Knight W.D. ``Manetic Resonance and relaxation'', Ed R. Blinc, North Holland 311 (1967)

\item D.E. Mac Laughlin ``Magnetic Resonance in the Superconducting state'' in Solid State Physics, Academic Press , New York 31 ,1 (1976) 

\item Moriya T., ``Nuclear Magnetic Relaxation in Antiferromagnetics'' Prog. Theor. Phys., 16, 23 (1956).

\item H. Niedoba, H. Launois, D. Brinkmann, R. Brugger and H. R. Zeller, ``NMR Study of the Low-Temperature Insulating State in the One-Dimensional ``Conductor'' $ K_2Pt (CN) 4Br0. 3.3H_2O$'',  Phys. Status Solidi, 58b 309(1973) DOI: 10.1002/pssb.2220580130

\item T. A. Platova, I. R. Mukhamedshin, H. Alloul, A. V. Dooglav and G. Collin, ``NQR and X-ray investigation of the structure of Na2/3CoO2 compound'',cond-mat/0908.0834,  Phys. Rev. B 80, 224106 (2009). DOI:10.1103/PhysRevB.80.224106

\item C. P. Slichter, ``Principles of Magnetic Resonance", Harper and Row (1963)(Springer-Verlag, New York, (1989), 3rd ed.

\item P. Wzietek, T. Mito, H. Alloul, D. Pontiroli, M. Aramini and M. Ricc\`o, ``NMR study of the Superconducting gap variation near the Mott transition in Cs3C60''  arXiv:1310.5529, Phys. Rev. Lett.  112, 066401 (2014).

\item Yosida K , ``Paramagnetic Susceptibility in Superconductors" Phys. Rev. 110 , 769 (1956)
\end{enumerate}

\section*{Further reading}

\begin{enumerate}
\item Abragam A. Bleaney B. ``Electronic Paramagnetic Resonance of Transition ions'' (Dover publications)(1986)

\item Alloul H. ``Introduction to the physics of Electrons in Solids '', Graduate texts in Physics, Springer (Heidelberg) (2011), ISBN 978-3-642-13564-4 DOI:10.1007/978-3-642-13565-1

\item Ashcroft Neil W., Mermin N. David, Saunders College (1976), ISBN 0030493463, 9780030493461

\item Kittel C., ``Introduction to Solid State Physics'', 8th Edition, Wiley (2005), ISBN 047141526X, 9780471415268

\item Tinkham M. , ``Introduction to Superconductivity'', Gordon and Breach (New York 1965), Dover Publications (1996), ISBN 0486134725, 9780486134727

\end{enumerate}
\section*{See also}

\begin{enumerate}

\item ``Bardeen-Cooper-Schrieffer theory'', Leon Cooper and Dimitri Feldman (2009), Scholarpedia, 4(1):6439.

\item ``Magnetic resonance imaging'', Joan Dawson and Paul C. Lauterbur Scholarpedia, 3(7):3381(2008)).

\item ``NMR in strongly correlated materials'', H. Alloul , Scholarpedia, 10(1):30632 (2015). 

\item \url{http://en.wikipedia.org/wiki/Nuclear_magnetic_resonance Nuclear magnetic resonance} in Wikipedia

\item \url{http://en.wikipedia.org/wiki/Peierls_transition Peierls transition} in Wikipedia

\end{enumerate}

\end{document}